\documentclass[twocolumn,showpacs,preprintnumbers,amsmath,amssymb]{revtex4}
 \usepackage{graphicx}
 \usepackage{dcolumn}
 \usepackage{bm}

\begin{document}
 \title{Non-adiabatic effects of superconductor silane under high pressure}
 \author{Fan Wei$^{1}$}
 \email{fan@theory.issp.ac.cn}
 \author{Wang Jiang-Long$^{2,1}$}
 \author{Zou Liang-Jian$^{1}$}
 \author{Zeng Zhi$^{1}$}
 \affiliation{$^{1}$Key Laboratory of Materials Physics, Institute of Solid
 State Physics, Hefei Institutes of Physical Sciences, Chinese
 Academy of Sciences, 230031-Hefei, People's Republic of China \\
 $^{2}$Department of Physics, Hebei University, 071002-Baoding, People's Republic of China}

\date{\today}

\begin{abstract}
Investigations of non-adiabatic effects by including vertex
corrections in the standard Eliashberg theory show that high phonon
frequency is unfavorable to superconductivity in regime of strong
vertex correction. This means that it is hard to find
high-transition-temperature superconductors in the compounds with
light elements if the non-adiabatic effects are strong. The interplay
interaction between non-adiabatic effect and Coulomb interaction makes
the transition temperature of silane superconductor not so high as
predicted by the standard Eliashberg theory.
\end{abstract}
 \maketitle

The recently found superconductor silane (SiH$_{4}$) at
high-pressure~\cite{Eremets} partially realizes the theoretical
predictions of high-temperature superconductors of hydrogen dominant
metallic alloy~\cite{Ashcroft1} and metallic hydrogen at high
pressure~\cite{Ashcroft2}. However, the transition temperature
($T_{c}$) is significantly lower than the values predicted in
previously theoretical calculations~\cite{Chen1,Feng1,Pickard1,Yao1}.
What is the underlaying mechanism of the lower $T_{c}$ compared with
the higher $T_{c}$ predicted in strong-coupling theory. With
increasing pressure the electronic structures of silane will evolve
from molecule-like energy-levels to electron-bands when gaseous silane
transforms into crystalline silane. The effective band-widths of
conducting electrons of metallic silane are narrow so that the
non-adiabatic effects will be prominent. It's very desirable study the
influences of non-adiabatic effects on $T_{c}$.

The non-adiabatic effects are partially included in the vertex
corrections in the theory of electron-phonon interaction. The
Eliashberg theory~\cite{Eliashberg1,Scalapino1,Allen1} combined with
the vertex corrections had been widely studied by perturbation
method~\cite{Kostur1,Grimaldi1,Fan1}. We have performed a full
parameter-space search based on Eliashberg theory with and without
vertex corrections~\cite{Fan1,Fan2}. In this paper, we study the
influences of the vertex corrections on $T_{c}$ of superconductor
silane at high pressure using reliable Eliashberg functions
$\alpha^{2}F(\omega)$ obtained from the calculations of
linear-response theory~\cite{Chen1}. Our results indicate that the
vertex correction can efficiently suppress $T_{c}$ approaching to the
values found in experiments. Additionally, we find that high phonon
frequency is unfavorable to high-$T_{c}$ if the strong vertex
corrections are included. This means that it is hard to realize
high-$T_{c}$ superconductors in silane and other hydrogen-rich
materials.

We have generalized the equation of energy gap in Ref.\cite{Kostur1}
by including the Coulomb interaction~\cite{Fan1}. The calculations of
vertex corrections are greatly simplified under isotropic
approximation. The electron-phonon interactions are included in the
vertex corrections only by the functions $\lambda_{n}$ defined as
$\lambda_{n}=2\int_{0}^{\infty}d\nu\alpha^{2}F(\nu)\nu/(\nu^{2}+\omega_{n}^{2})$.
When temperature is very close to $T_{c}$, the energy-gap equation
including the leading vertex correction from electron-phonon
interaction is written as
$\sum_{n'=-\infty}^{+\infty}K_{nn'}\Delta_{n'}/|\omega_{n'}|=0$ with
the kernel matrix
\begin{equation}\label{GapKN}
 K_{nn'}=[\lambda_{n-n'}B_{nn'}-\mu^{*}+C_{nn'}]a_{n'}-\delta_{nn'}H_{n'}
 \end{equation}
\noindent where the definitions of $H_{n}$, $A_{nn'}$, $B_{nn'}$,
$C_{nn'}$ and $a_{n}$ are presented in Ref.\cite{Kostur1,Fan1}. In
order to calculate $T_{c}$, the matrix $K_{nn'}$ is symmetrized with
the same manner as in Ref.\cite{Allen1}. The $E_{B}$ is the effective
band-width and the Coulomb pseudo-potential $\mu^{*}$ is defined as
$\mu^{*}=\mu_{0}/(1+\mu_{0}\ln(E_{B}/\Omega_{0}))$, where
$\mu_{0}=N(0)U$, $N(0)$ the density of state at Fermi energy $E_{F}$,
U the Coulomb interaction between electrons and $\Omega_{0}$
characteristic energy of typical phonon correlated to
superconductivity. Generally, the ratio $\Omega_{0}/E_{B}$ takes as
the parameter to measure vertex correction and the larger
$\Omega_{0}/E_{B}$ is corresponding to the stronger vertex correction.
In this context, the vertex corrections are controlled by $E_{B}$, the
smaller $E_{B}$ (or larger $\Omega_{0}/E_{B}$) for stronger vertex
corrections. With development of method of electronic-structure
calculation in solid materials, the parameters of electron-phonon
interaction can be calculated using density functional theory combined
with perturbing linear-response theory~\cite{Savrasov1,Baroni1}. The
Elaishberg functions $\alpha^{2}F(\omega)$ obtained by linear response
theory are easily merged in the formulae used in this context.

\begin{figure}[h]
\includegraphics[width=0.45\textwidth]{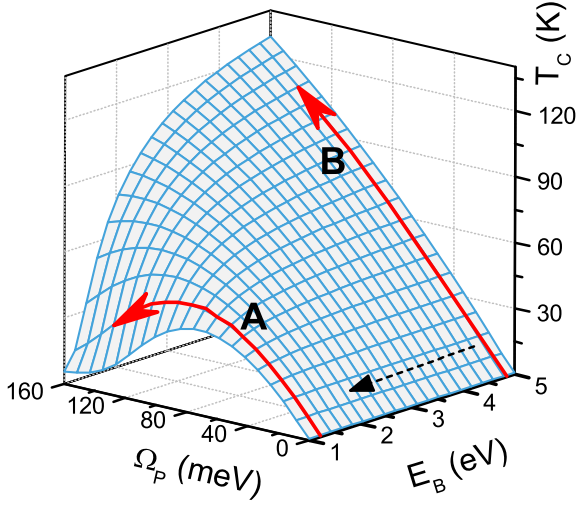}
\caption{\label{fig1} The $T_{c}$ on $\Omega_{P}-E_{B}$ parameter
plane. The two solid arrows {\bf A} and {\bf B} show the changes of
$T_{c}$ with $\Omega_{P}$ at fixed $E_{B}$ ({\bf A}) near 1 eV and
({\bf B})5 eV respectively. The dashed arrow shows the direction
enhanced the vertex corrections.}
\end{figure}

\begin{figure}[h]
\includegraphics[width=0.45\textwidth]{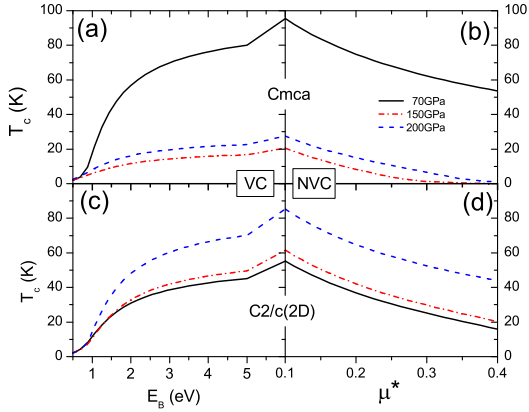}
\caption{\label{fig2} The dependence of $T_{c}$ on $E_{B}$ and
$\mu^{*}$, based on the Eliashberg functions $\alpha^{2}F(\omega)$
obtained from linear-response theory at different pressures for the
$Cmca$ structure (a,b) and $C2/c$(2D) structure (c,d)
respectively~\cite{Chen1}. The (a) and (c) include vertex corrections
(VC), and there are no vertex corrections (NVC) in (b) and (d).}
\end{figure}

In the standard Eliashberg theory, $T_{c}$ will infinitely increase
with phonon frequency or energy. It's widely expected that
high-temperature superconductors should be found in compounds
containing light elements such as hydrogen-rich material silane and
metal hydrogen at high pressure. We will see that the non-adiabatic
effects set bounds to infinitely increasing $T_{c}$. In order to
obtain very general results, at first step, we apply the simple phonon
spectrum and Eliashberg function $\alpha^{2}F(\omega)$ used in
Ref.\cite{Scalapino1}. The Fig.\ref{fig1} shows how $T_{c}$ changes
with $\Omega_{P}$ and $E_{B}$. We can see that for large band-width
$E_{B}$=5 eV (weak vertex correction), $T_{c}$ increases monotonously
with $\Omega_{P}$ just as in the standard Eliashberg theory. However
for very strong vertex correction with small band-width $E_{B}$=1 eV,
$T_{c}$ is non-monotonously dependent on $\Omega_{P}$. When
$\Omega_{P}$ is larger than a threshold value, $T_{c}$ will decrease
with $\Omega_{P}$. This means that, if the vertex corrections are
included, high phonon energy is unfavorable to superconductivity.

We preform the calculations of the vertex corrections of real
superconducting material silane containing the lightest element:
hydrogen. The lowest order vertex correction can significantly reduce
$T_{c}$ when energy of phonon $\Omega_{P}$ is larger than 80 meV shown
in Fig.\ref{fig1}. We adopt the Eliashberg functions that had been
reported in Ref~\cite{Chen1} in the calculations of linear-response
theory. The structures of crystal silane at high-pressures haven't
been completely defined. The metallic $P6_{3}$ structure had been
found in experiment~\cite{Eremets}. The theoretical layered structure
with $Cmca$ space-group symmetry is more stable from 60-200
GPa~\cite{Chen1}, other stable structures at higher pressures had been
already reported~\cite{Canales1}. The hydrogen-rich superconductor
silane at higher pressure has very high phonon energies coming from
the vibrations of hydrogen atoms. The effective phonon energies
$\langle\omega\rangle_{ln}$ for silane at high pressures distribute
from 50 meV to 75 meV dependent on pressures. The parameters $\lambda$
of electron-phonon interaction are about 1.17, 0.62, 0.75 for the
$Cmca$ structure and 0.84, 0.87, 1.1 for the $C2/c$(2D) structure
respectively when pressure increases from 70 GPa to 150 GPa and to 200
GPa. The $T_{c}$ calculated using Allen-McMillan formula are
distributed from 20 K to 80 K~\cite{Chen1} that are greatly larger
than experimental $T_{c}<17$K.

\begin{figure}[h]
 \includegraphics[width=0.45\textwidth]{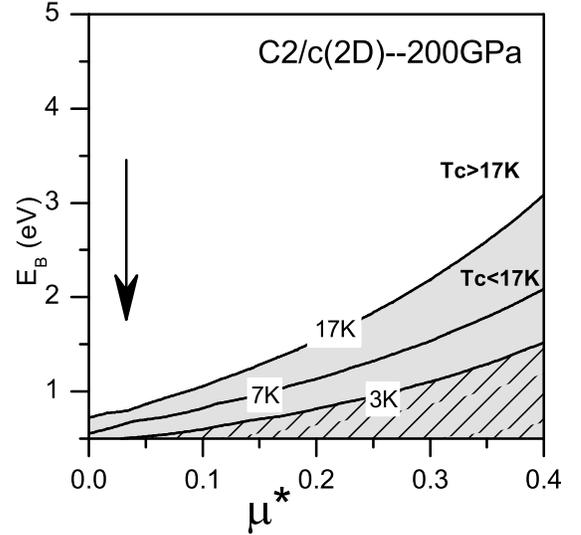}
 \caption{\label{fig3} The $T_{c}$ on $\mu^{*}$-$E_{B}$ plane
 using the Eliashberg functions $\alpha^{2}F(\omega)$ obtained from
 linear-response theory~\cite{Chen1}. The structure has $C2/c$(2D) space
 group symmetry at pressure 200 GPa.  The shade areas are regions
 with $T_{c}$$<$17 K. The arrow shows the directions of increasing
 the vertex correction.}
\end{figure}

As reported in Ref.\cite{Chen1}, the pressure-dependence $T_{c}$ of
the $Cmca$ structure is different from the $C2/c$(2D) structure. The
$T_{c}$ of the $Cmca$ structure decreases when pressure increases from
70 GPa to 150 GPa, and slightly increases with increasing pressure
further to 200 GPa. However for the $C2/c$(2D) structure, $T_{c}$
monotonously increases with pressures from 70 GPa to 200 GPa. We first
present the results of the $Cmca$ structure in detail shown in
Fig.\ref{fig2}(a,b). If the vertex corrections are not included and
$\mu^{*}$=0.1, the values of $T_{c}$ are close to 20 K at pressures
150 GPa and 200 GPa, which are smaller than those obtained from
McMillan formula. So our results indicate that for very broad
distributed phonon spectrum, the standard strong-coupling calculations
are needed to accurately calculate $T_{c}$ beyond simple McMillan and
Allen $T_{c}$ formulae.

At low pressure with larger inter-molecule distances, the crystalline
silane is an insulator with narrow energy bands corresponding to
discrete energy levels of a single molecule. The effective band-widths
will increase with increasing pressures and overlaps of molecular
orbits of different molecules. If the vertex corrections are included
in our calculations with the effective band-width $E_{B}$ of
conducting electrons from 0.5 eV to 5 eV, the $T_{c}$ at 150 GPa and
200 GPa are significantly reduced to the value smaller than 17 K shown
in Fig.\ref{fig2}(a). The effective band-widths obtained from the
density functional calculations increase from very small values at low
pressure to about 1.0-3.0 eV at 100-200 GPa for the $C2/c$(2D) and
Cmca structures~\cite{Chen1}. In another case, the effects of Coulomb
interaction are more significant because the $T_{c}$ decreases to very
small values with increasing $\mu^{*}$ at 150 GPa and 200 GPa shown in
Fig.\ref{fig2}(b). If we want to know what is more important for the
depression of $T_{c}$ we should know real value of $\mu^{*}$ at 150
GPa and 200 GPa. At 70 GPa, the small band-widths with $E_{B}$$<$1 eV
guarantee the $T_{c}$ lower than 17 K. We find that, if without vertex
correction, even larger $\mu^{*}$ (=0.4) can not suppress $T_{c}$ to
value smaller than 17 K at 70 GPa.

The effects of vertex correction of the $C2/c$(2D) structure in
Fig.\ref{fig2}(c) are similar to the $Cmca$ structure in
Fig.\ref{fig2}(a), however the influences of the Coulomb interaction
in Fig.\ref{fig2}(d) are different from the $Cmca$ in
Fig.\ref{fig2}(b). The $C2/c$(2D) structure is less stable than the
$Cmca$ structure. The stronger electron-phonon interaction induces its
structure more unstable at high pressures from 150 GPa to 200 GPa.
Theoretical values of $T_{c}$ increase with pressures. This is because
the enhancements of $T_{c}$ induced by the increase of $\lambda$ are
larger than the depressions of $T_{c}$ due to the vertex correction.
For the $C2/c$(2D) structure, the Coulomb interaction depresses
$T_{c}$ to values which are still larger than 17 K shown in
Fig.\ref{fig2}(b). So Coulomb interaction individually can not explain
why $T_{c}$ smaller than 17 K in high-pressure experiments. The
effective band-widths $E_{B}$ of conducting electrons in the
$C2/c$(2D) structure are larger than 1 eV at 150 GPa and 200
GPa~\cite{Chen1} so that the vertex correction individually can not
explain low $T_{c}$ at 100 GPa and 200 GPa as well as shown in
Fig.\ref{fig2}(c). The Fig.\ref{fig3} illustrates the $T_{c}$ for the
$C2/c$(2D) structure on $\mu^{*}$-$E_{B}$ plane at pressure 200 GPa.
The shading region in the figure shows the effective region with
parameters $\mu^{*}$ and $E_{B}$ that can explain the $T_{C}$ smaller
than 17 K, which is only small area of parameter-space with
$\mu^{*}$$>$0.1 and $E_{B}$$<$2.0 eV. Especially, the hatched region
shows the parameter-space to explain why experimental $T_{c}$ is
smaller 3 K at 200 GPa. So our results indicate that the interplay of
vertex correction and Coulomb interaction can explain the low $T_{c}$
of the C2/c(2D) structure. Our calculations can not fully explain
pressure-dependent $T_{c}$ in Ref~\cite{Eremets}, especially, the
higher $T_{c}$=17 K near 100-120 GPa because of the lack of
information of structural changes.

In summary, we study the effects of the vertex correction on $T_{c}$
for superconductor of silane at high pressure. Our results indicate
that the non-adiabatic effects are the barrier to prevent $T_{c}$ from
increasing infinitely with phonon frequency. This means that it is
hard to realize high $T_{c}$ to home temperature in high pressure
rich-hydrogen materials. Our results also show that the interplay
interaction between Coulomb interaction and vertex correction is
essentially to explain the larger differences between theory and
experiments.

The author (Fan W) thanks Dr. Li Yan-Ling for the helpful discussions.
The work preforms in the Center for Computational Science of CASHIPS
and is supported by Knowledge Innovation Project of Chinese Academy of
Science.


\begin{thebibliography}{99}

 \bibitem{Eremets} Eremets M I, Medvedev I A, Tse J S and Yao Y 2008 Science {\bf 319} 1506
 \bibitem{Ashcroft1} Ashcroft N W 2004 Phys. Rev. Lett. {\bf 92} 187002
 \bibitem{Ashcroft2} Ashcroft N W 1968 Phys. Rev. Lett. {\bf 21} 1748
 \bibitem{Chen1} Chen X J, Wang J L, Struzhkin V V, Mao H K, Hemley R J and Lin H Q 2008
 Phys. Rev. Lett. {\bf 101} 077002 and supplemental materials
 \bibitem{Feng1} Feng J, Grochala W, Jaro\'n T, Hoffmann R, Bergara A
 and Ashcroft N W 2006 Phys. Rev. Lett. {\bf 96} 017006
 \bibitem{Pickard1} Pickard C J and Needs R J 2006 Phys. Rev. Lett. {\bf 97} 045504
 \bibitem{Yao1} Yao Y, Tse J S, Ma Y and Tanaka K 2007 Eur. Phys. Lett. {\bf 78} 37003
 \bibitem{Eliashberg1} \'Eliashberg G M 1960 Soviet. Phys. JETP, {\bf 11} 696
 \bibitem{Scalapino1} Scalapino D J, Schrieffer J R and Wilkins J W 1966 Phys. Rev. {\bf 148} 263
 \bibitem{Allen1} Allen P B and Dynes R C 1975 Phys. Rev. {\bf B 12} 905
 \bibitem{Kostur1} Kostur V N and Mitrovi\'c B 1994 Phys. Rev. {\bf B 50} 12774
 \bibitem{Grimaldi1} Grimaldi C, Pietronero L and Str\"assler S 1995 Phys. Rev. {\bf B 52} 10530
 \bibitem{Fan1} Fan W 2009 Physica C {\bf 469} 177
 \bibitem{Fan2} Fan W 2008 Chinese Phys. Lett. {\bf 25} 2217
 \bibitem{Savrasov1} Savrasov S Y and Savrasov D Y 1996 Physical Review {\bf B 54} 16487
 \bibitem{Baroni1} Baroni S, de Gironcoli S, Dal Corso A and Giannozzi P 2001 Rev. Mod. Phys. {\bf 73} 515
 \bibitem{Canales1} Martinez-Canales M, Oganov A R, Ma Y M, Yan Y, Lyakhov A O
 and Bergara A 2009 Phys. Rev. Lett. {\bf 102} 087005
 \end{thebibliography}
\end{document}